\documentclass[%
reprint,
superscriptaddress,
 aip, 
 amsmath,amssymb,
 aps,
]{revtex4-2}

\usepackage{graphicx}
\usepackage{dcolumn}
\usepackage{bm}
\begin{document}


\bibliographystyle{unsrt}

\title{Effect of non-local transport of hot electrons on the laser-target ablation}

\author{Z. H. Chen}
\affiliation{Department of Nuclear Science and Technology, National University of Defense Technology,\\Changsha 410073, China}%
 
\author{X. H. Yang}\thanks{xhyang@nudt.edu.cn}
\affiliation{Department of Nuclear Science and Technology, National University of Defense Technology,\\Changsha 410073, China}%
\affiliation{Collaborative Innovation Centre of IFSA, Shanghai Jiao Tong University, Shanghai, 200240, China}%

\author{G. B. Zhang}
\affiliation{Department of Nuclear Science and Technology, National University of Defense Technology,\\Changsha 410073, China}%

\author{Y. Y. Ma}\thanks{yanyunma@126.com}
\affiliation{Collaborative Innovation Centre of IFSA, Shanghai Jiao Tong University, Shanghai, 200240, China}%
\affiliation{College of Advanced Interdisciplinary Studies, National University of Defense Technology, Changsha 410073, China}%

\author{H. Xu}
\affiliation{Collaborative Innovation Centre of IFSA, Shanghai Jiao Tong University, Shanghai, 200240, China}%
\affiliation{College of Computing Science, National University of Defense Technology, Changsha 410073, China}%

\author{S. X. Luan}\thanks{sxluan@siom.ac.cn}
\affiliation{State Key Laboratory of High Field Laser Physics and CAS Center for Excellence in Ultra-Intense Laser Science, Shanghai Institute of Optics and Fine Mechanics (SIOM), Chinese Academy of Sciences (CAS), Shanghai 201800, China}%

\author{J. Zhang}
\affiliation{Collaborative Innovation Centre of IFSA, Shanghai Jiao Tong University, Shanghai, 200240, China}%
\affiliation{Key Laboratory for Laser Plasmas (Ministry of Education), School of Physics and Astronomy,\\Shanghai Jiao Tong University, Shanghai, 200240, China}%

\begin{abstract}
The non-local heat transport of hot electrons during high-intensity lasers interaction with plasmas can preheat the fuel and limit the heat flow in inertial confinement fusion. It increases the entropy of the fuel and decreases the final compression. In this paper, the non-local electron transport model that is based on the improved SNB algorithm has been embedded into the radiation hydrodynamic code and is benchmarked with two classical non-local transport cases. Then we studied a 2$\omega$ laser ablating a CH target by using the non-local module. It is found that the non-local effect becomes significant when the laser intensity is above $1\times 10^{14} \mathrm{W/cm^{2}} $. The mass ablation rate from the SNB model is increased compared to that of the flux-limited model due to the lower coronal plasma temperature. This non-local model has a better agreement with the experimental results compared to that of the flux-limited model. The non-local transport is strongly dependent on the laser frequency, and the thresholds that the non-local transport should be considered are obtained for lasers of different frequencies. The appropriate flux-limiters that should be employed in the flux-limited model for different lasers are also presented. The results here should have a good reference for the laser-target ablation applications.

\end{abstract}

\maketitle


\section{INTRODUCTION}
The theory of heat conduction in plasmas with Coulomb collisions was derived in the 1950s, i.e., the Spitzer-Härm (SH) theory \cite{spitzerTransportPhenomenaCompletely1953}. The SH theory is applicable for the cases with the mean free path $\lambda_{e}$ of the electron is much smaller than the temperature gradient scale length $L\left( L=T/ \nabla T \right )$, where the heat conduction is driven by collisions in a local area. The SH theory is suitable for the condition of the Knudsen number\cite{moraNonlocalElectronTransport1994} $\left ( {\rm{Kn}} = {\lambda _e}/L \right ) $ smaller than $2\times 10^{-3}$.

However, a steep temperature gradient usually exists in inertial confinement fusion (ICF). The mean free path $\lambda_{e}$ of such hot electrons no longer satisfies the condition of which much smaller than the gradient scale length of the temperature. And the transport of hot electrons cannot be described correctly by the classical theory. The electron distribution function (EDF) will deviate from the Maxwellian distribution, producing a high energy tail. The tail electrons have a long mean free path and can deposit their energy into the inner shell by non-local transport. And the heat carrying electrons ($v\approx 3.7v_{Te}$, $v_{Te}$ is the electron thermal velocity) are relatively reduced, which will decrease the heat flow around the critical surface. As the power of the laser increases, the laser plasma instability (LPI) like stimulated Raman scattering (SRS), stimulated Brillouin scattering (SBS), and two plasmon decay (TPD) will occur in the plasma. Hot electrons are produced and their distribution will further deviate from the Maxwellian distribution\cite{milderImpactNonMaxwellianElectron2019,milderMeasurementsNonMaxwellianElectron2021,brantovNonlocalTransportHot2014,brantovNonlocalTransportHot2013}. It changes the laser absorptivity and ablation rate, preheating the internal fuel and reducing the final compression \cite{shvartsRoleFastelectronPreheating2008,christophersonDirectMeasurementsDT2021}.

In conventional hydrodynamic simulations, a flux-limiter is used to limit the heat flow to match experimental results which are much smaller than that predicted by the SH theory. The maximum value of heat flow in plasmas is the free heat flow ${q_{\max ,e}} = {n_e}{k_B}{T_{e{\rm{ }}}}\sqrt {{{{k_B}{T_e}} \mathord{\left/
{\vphantom {{{k_B}{T_e}} {{m_e}}}} \right.\kern-\nulldelimiterspace} {{m_e}}}} $, where  $n_{e}$ is the electron density, $k_{B}$ is the Boltzmann constant, $T_{e}$ is the electron temperature, and $m_{e}$ is the electron mass. The flux-limited model combines a flux-limiter $f$ times the free heat flow $q_{max}$ and the SH heat flow $q_{0}$ at steep temperatures to obtain the modified heat flow $q$. One of the simplest models is to take  $q=min\left ( q_{0},fq_{max} \right ) $. For planar targets\cite{goncharovAblativeRichtmyerMeshkov2006}, the value of $f$ is usually taken to be between $0.03\sim 0.1$, while for spherical targets, the flux-limiter $f$ of $0.06\sim 0.5$ is more appropriate \cite{brantovNonlocalTransportHot2013}. For a specific target, this exact flux-limiter $f$ is usually not based on physics, but estimated from the summary of relevant experiments. Therefore, the value of $f$ tends to be different for different experiments. The results of the simulations and the experiments do not match exactly, even though the flux-limiter is employed. Moreover, the flux-limited model does not consider the effect of hot electron preheating, which will be affected by the laser power and wavelength \cite{huValidationThermalTransportModeling2008}.

For some longer wavelength lasers, such as a $1\omega\ \left(1.06\ \rm{\mu m}\right)$ laser, the energy absorption efficiency is relatively low, and only 30\% to 40\% for intensity between $10^{13}\ \mathrm{W/cm^{2}}$ and $10^{15}\  \mathrm{W/cm^{2}}$, while the energy absorptivity of $0.53\ \mathrm{\mu m}\left(2\omega\right)$ laser can reach 60\% to 80\%\cite{garban-labauneEffectLaserWavelength1982}. Higher energy absorptivity for the shorter laser wavelength, $2\omega$ and $3\omega \left(0.351\ \mathrm{\mu m}\right)$, have advantages in inertial confinement fusion. The $3\omega$ laser is currently the widely used laser in ICF due to the higher energy absorptivity and lower level of LPI that scales to $I\lambda^{2}$, where $I$ is the laser intensity, $\lambda$ is the laser wavelength. Therefore, shorter wavelength lasers are usually used to reduce the hot electrons generated by LPI and thus preheating. However, the $3\omega$ laser cause more damage risk to the optical elements, and the allowed laser pulse bandwidth is narrow, inducing a limitation of the laser energy\cite{glenzerLaserMatterInteractions5272007}. The $2\omega$ laser is a compromise choice, which has a higher damage threshold for optical elements than that of the $3\omega$ laser. In the NIF, the maximum energy of a $2\omega$ laser is 1.5 times higher than that of $3\omega$ laser, which can deliver more energy to the target\cite{suterProspectsHighgainHigh2005,mosesNationalIgnitionFacility2008}. In addition, it is easier to control the green light $\left(2\omega\right)$ propagation in the cavity. In summary, the $2\omega$ laser may be more suitable than the UV laser $\left(3\omega\right)$ for direct driving compression \cite{zhangDoubleconeIgnitionScheme2020,wuMachinelearningGuidedOptimization2022,liDesignLaserPulse2022} in specific situations. However, the development of non-local heating of fast electrons generated by the $2\omega$ laser and its influence on ablation process are still unclear.

The Vlasov-Fokker-Planck (VFP) kinetic model is usually used to simulate the non-local heat transport conduction\cite{brodrickTestingNonlocalModels2017,holecNonlocalTransportHydrodynamic2018,sherlockComparisonNonlocalElectron2017}. However, it is computationally expensive to extend to high dimensions and difficult to couple other physical effects in inertial confinement fusion. Thus, an approximate model that can accurately simulate the non-local conduction of electrons is preferred. Many non-local models have been developed, such as the Schurtz-Nicolaï-Busquet (SNB) model \cite{schurtzNonlocalElectronConduction2000,caoImprovedNonlocalElectron2015}, the Colombant-Manheimer-Goncharov (CMG) model \cite{manheimerDevelopmentKrookModel2008}, and various models improved by Holec \cite{holecNonlocalTransportHydrodynamic2018}, Sijoy \cite{sijoyImprovedFullyImplicit2017} and Chrisment \cite{chrismentAnalysisKineticModel2022}, respectively. The SNB multigroup diffusion model is one of the widely used models, which is much more accurate and easily expand to multi-dimensions. With the different collision operators, the SNB model can be divided into BGK-SNB \cite{schurtzNonlocalElectronConduction2000,hunanaGeneralizedFluidModels2022} and AWBS-SNB \cite{delsorboReducedEntropicModel2015} models. The BGK operator is of the following form: $C\left( {{f_0}} \right) =  - {\nu _{ee}}\left( {{f_0} - f_0^m} \right)$, while the AWBS operator is $C\left( {{f_0}} \right) = {\nu _{ee}}v{\partial _v}\left( {{f_0} - f_0^m} \right)$, where $\nu_{ee}$ is the electron-electron collision frequency, $f_{0}$ is the 0th order correction to the electron distribution function, $f_{0}^{m}$ is Maxwellian EDF, and $v$ is the electron velocity. Compared to the BGK operator, the AWBS operator allows diffusion between energy groups, but is not easily implemented in hydrodynamic programs. So, the implicit BGK-SNB model proposed by Cao et al. \cite{caoImprovedNonlocalElectron2015} is adopted and applied to the radiation hydrodynamic code.

In this paper, we briefly describe the SNB model in Section 2. In Section 3, we benchmark the non-local module added in the FLASH code \cite{fryxellFLASHAdaptiveMesh2000,meineckeStrongSuppressionHeat2022} with two simple temperature transport cases. In Sections 4 and 5 the high-intensity laser irradiation on a target is studied. The non-local effect of lasers of different intensities and wavelengths are investigated. Finally, a conclusion is presented in Section 6.

\section{MODEL OF THE NON-LOCAL HEAT TRANSPORT}
In order to accurately simulate the non-local heat conduction of electrons in plasma, we added the SNB model to the FLASH code. The FLASH code is a publicly available, parallel finite-volume Eulerian mesh program with adaptive mesh refinement. It includes varies physical processes, like heat conduction, heat exchange, and multigroup radiative transfer. It is capable of modeling 3T radiative hydrodynamics and is widely used in the areas of astrophysics and high-energy density physics.

The SNB model is a non-local heat conduction model with multigroup diffusion \cite{schurtzNonlocalElectronConduction2000,nicolaiPracticalNonlocalModel2006}. The multigroup diffusion equation for the non-local heat flow is
\begin{equation} \label{Eq1}
\left[ {\frac{{\rm{r}}}{{{\lambda _g}({\bf{r}})}} - \nabla  \cdot \frac{{\lambda _g^E({\bf{r}})}}{3}\nabla } \right]{H_g}({\bf{r}}) =  - \nabla  \cdot {{\bf{U}}_g}({\bf{r}}),
\end{equation}
where $\mathbf{U} _{g}$ is the heat flow for different energy groups, $\lambda_{g}$ is the mean free path of electrons for different energy groups, and $\lambda_{g}^{E}$ denotes the mean free path considering the local electric field correction, and r is a dimensionless number. The expressions are respectively as follows:
\begin{equation} \label{Eq2}
{{\bf{U}}_g} = \frac{{\kappa \nabla {T_e}}}{{24}}\int_{{E_{g - 1}}/{k_B}{T_e}}^{{E_g}/{k_B}{T_e}} {{\beta ^4}{e^{ - \beta }}d\beta },
\end{equation}
\begin{equation} \label{Eq3}
{\lambda _g} = 2{\left( {{E_{g - 1/2}}/{k_B}{T_e}} \right)^2}{\lambda _e},
\end{equation}
\begin{equation} \label{Eq4}
\frac{1}{{\lambda _{\rm{g}}^E}} = \frac{1}{{{\lambda _g}}} + \frac{{\left| {\boldsymbol{\epsilon }} \right|}}{{{E_{g - 1/2}}}},
\end{equation}
where $\kappa$ is thermal conductivity, $E_{g}$ is the upper energy boundary of the electron energy group $g$, $k_{B}$ is Boltzmann constant and $\boldsymbol {\epsilon}$ is the local electric field
\begin{equation} \label{Eq5}
{\boldsymbol{\epsilon }} = {k_B}{T_e}\left( {\nabla \ln \left( {{n_e}} \right) + \gamma \nabla \ln \left( {{T_e}} \right)} \right),
\end{equation}
where $\gamma$ is a function of the average ionization degree $\bar{Z}$, $\gamma \left( {\bar{Z}} \right) = 1 + \left( {1.5\bar{Z} + 0.715} \right)/\left( {\bar{Z} + 2.15} \right)$.

For the original SNB model, the non-local heat flux is solved explicitly, which requires the time step should be smaller than $10^{-15}$s. An alternative approach is to solve the heat conduction equation implicitly by calculating the effective thermal conductivity $K_{eff}=\frac{\mathbf{Q}_{nl}}{\nabla T_e}$, where $\mathbf{Q}_{nl}=\mathbf{Q}_{SH}-\sum_{g}{\frac{\lambda_g}{3}\nabla H_g}$ is the non-local heat flux.. However, it is difficult to deal with the preheat region and corona. $K_{eff}$ may have a singularity in the preheat region due to the temperature gradient is zero. On the other hand, it may become negative in the corona due to the non-local heat flux is opposite to the temperature gradient. Therefore, using the effective thermal conductivity is not a reliable method. Duc Cao improved the SNB algorithm by calculating the divergence of the non-local heat flux ($\nabla  \cdot {{\bf{Q}}_{nl}} =  - \sum\limits_g {{H_g}/{\lambda _g}}$) implicitly instead of explicit solution\cite{caoImprovedNonlocalElectron2015}. The improved implicit SNB algorithm relaxes the time step while ensuring the accuracy. According to the implicit iterative method proposed by Cao et al., the thermal conduction equation is rewritten to
\begin{equation} \label{Eq6}
\rho {c_v}\frac{{\Delta {T_e}}}{{\Delta t}} = \nabla  \cdot {{\bf{Q}}_{nl}},
\end{equation}
where $c_{v}$ is the specific heat at constant volume. The following convergence condition is set
\begin{equation} \label{Eq7}
\left| {\nabla  \cdot \kappa _{SH}^n\nabla T_e^{k + 1} - \nabla  \cdot \kappa _{SH}^n\nabla T_e^k} \right| \le \alpha \rho {c_v}\frac{{T_e^k}}{{\Delta t}}
\end{equation}
where $\alpha=0.01$ is an adjustable convergence factor. If the convergence condition is satisfied in all regions of interest, set $T_e^{n + 1} = T_e^{k + 1}$, and jump out of the loop. Otherwise, solve Eq. (\ref{Eq1}) using $T^{k+1}$ and re-calculate $\nabla  \cdot {\bf{Q}}_{nl}^{k + 1}$ for the next iteration.

It is worth noting that the upper bound of the hot electron energy group needs to be large enough to ensure that the multigroup diffusion equation correctly reflect the effect of non-local transport on hot electrons. The upper bound $E_{g,max}$ can be typically taken to be 15 times the maximum value of the electron temperature $k_{B}T_{e,max}$, such that the integral $\int_0^{15} {{\beta ^4}{e^{ - \beta }}d\beta  \approx {\rm{23}}{\rm{.9794}}} $ in $\bf{U_{g}}$ is sufficiently close to its maximum value 24. On the other hand, there are discrepancies in the dimensionless number r in various studies. We use the modified mean free path proposed by Brodrick et al. \cite{brodrickTestingNonlocalModels2017,sherlockComparisonNonlocalElectron2017}, where the dimensionless number r = 2 and $\lambda_g=2{\sqrt2\left(\frac{E_g}{k_BT_e}\right)}^2\lambda_{ei}$, instead of r = 4 in the original paper\cite{schurtzNonlocalElectronConduction2000}. Compared with the original mean free path, the modified mean free path allows the SNB code and the VFP code to maintain good agreement for different materials.

\section{BENCHMARK OF THE MODEL}
\begin{figure}[htbp]\centering
	\includegraphics[width=6cm]{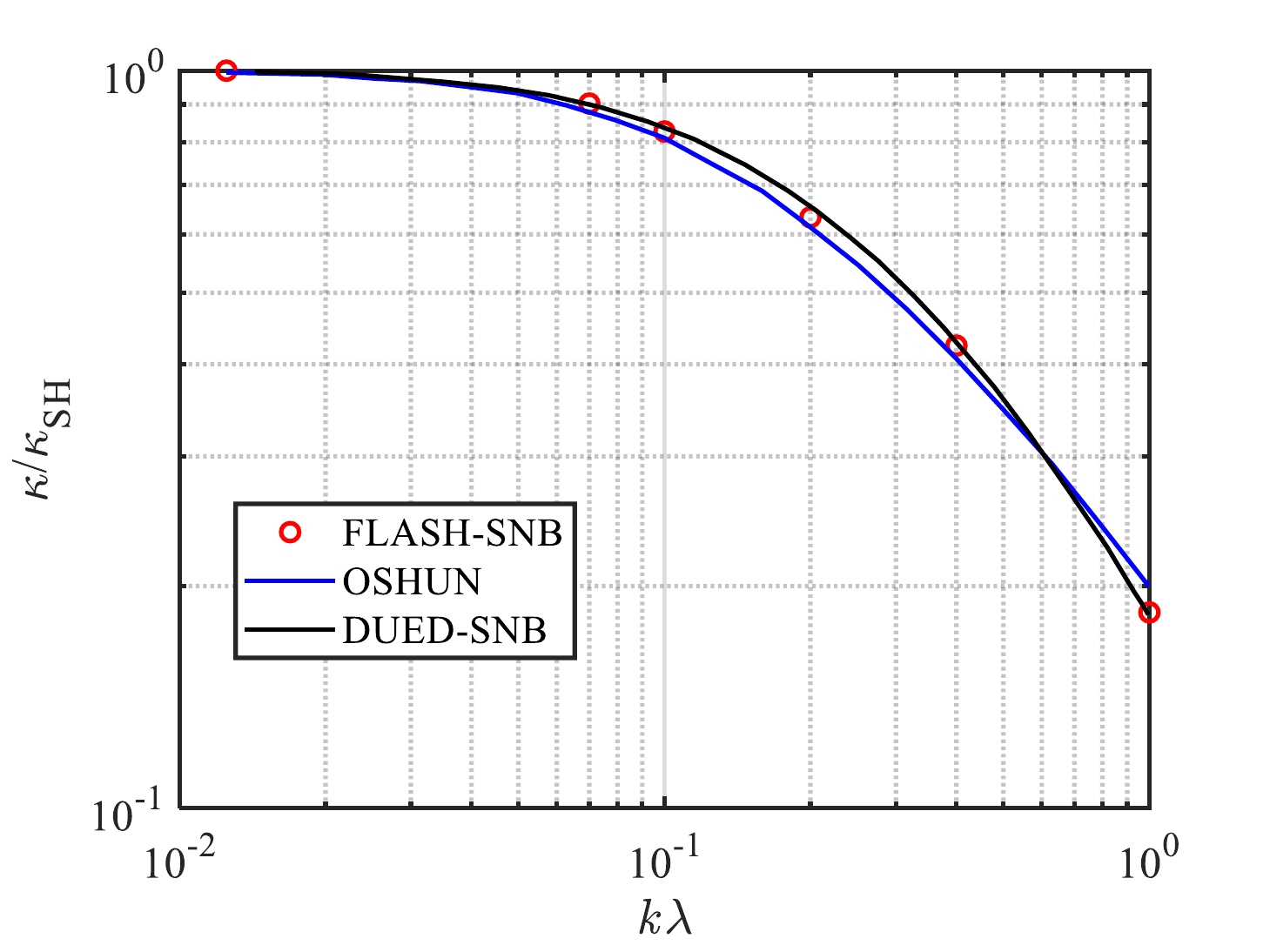}
	\caption{\label{fig1} The ratio of effective thermal conductivity to SH thermal conductivity varies with $k\lambda$ for $Z=1$, where $k$ is the wave number and $\lambda$ is the electron mean free path. The red dots are the results obtained from our modified FLASH-SNB code, the blue curve (OSHUN) and the black curve (DUED code) are both from the paper of Marocchino.}
\end{figure}

Firstly, we validate the non-local module with a classic example, i.e., the Epperlein-Short test \cite{epperleinPracticalNonlocalModel1991}. Assuming the temperature has a small perturbation, the heat conduction equation $\rho {c_v}{\partial _t}{T_e} = \nabla  \cdot \kappa \nabla {T_e}$ has a close-form solution ${T_e} = {T_{e,0}} + {T_{e,1}}{e^{ - \gamma }}\cos (kx)$, where $T_{e,0}$ is the initial temperature, $T_{e,1}$ is the perturbation temperature, $T_{e,1}\ll T_{e,0}$ is the applicability condition, $\gamma  = {{{k^2}t\kappa } \mathord{\left/{\vphantom {{{k^2}t\kappa } {\rho {c_v}}}} \right.\kern-\nulldelimiterspace} {\rho {c_v}}}$ is the decay rate. Comparing the decay rate $\gamma$ obtained from the non-local SNB model with the result $\gamma_{SH}$ from the classical SH model, the ratio of the thermal conductivity between the two cases is obtained, as shown in Fig. \ref{fig1}. The simulation uses the same parameters as Marocchino et al. \cite{marocchinoComparisonNonlocalHydrodynamic2013}. The background is a homogeneous H plasma with $Z=1$, density $n_{e}=1\times10^{23}\rm{cm^{-3}}$, initial temperature $T_{e,0}=307\ \rm{eV}$, and the perturbation temperature $T_{e,1}=12.5\ \rm{eV}$. The electron mean free path is ${\lambda _e} = {{3{{\left( {{k_B}T} \right)}^2}} \mathord{\left/
{\vphantom {{3{{\left( {{k_B}T} \right)}^2}} {\left( {4\sqrt {2\pi } {e^4}Z{n_e}\ln \Lambda } \right)}}} \right.\kern-\nulldelimiterspace} {\left( {4\sqrt {2\pi } {e^4}Z{n_e}\ln \Lambda } \right)}}$. The results are shown in Fig. \ref{fig1}, and ${\kappa  \mathord{\left/{\vphantom {\kappa  {{\kappa _{sh}}}}} \right.
\kern-\nulldelimiterspace} {{\kappa _{SH}}}}$ is obtained by calculating the decay rate of SH and SNB models at $x = 0$. It can be seen that the results of FLASH-SNB are in good agreement with that of the OSHUN (a Fokker-Planck code) and DUED-SNB given by Marocchino \cite{marocchinoComparisonNonlocalHydrodynamic2013}.

\begin{figure}[htbp]\centering\suppressfloats
	\includegraphics[width=6cm]{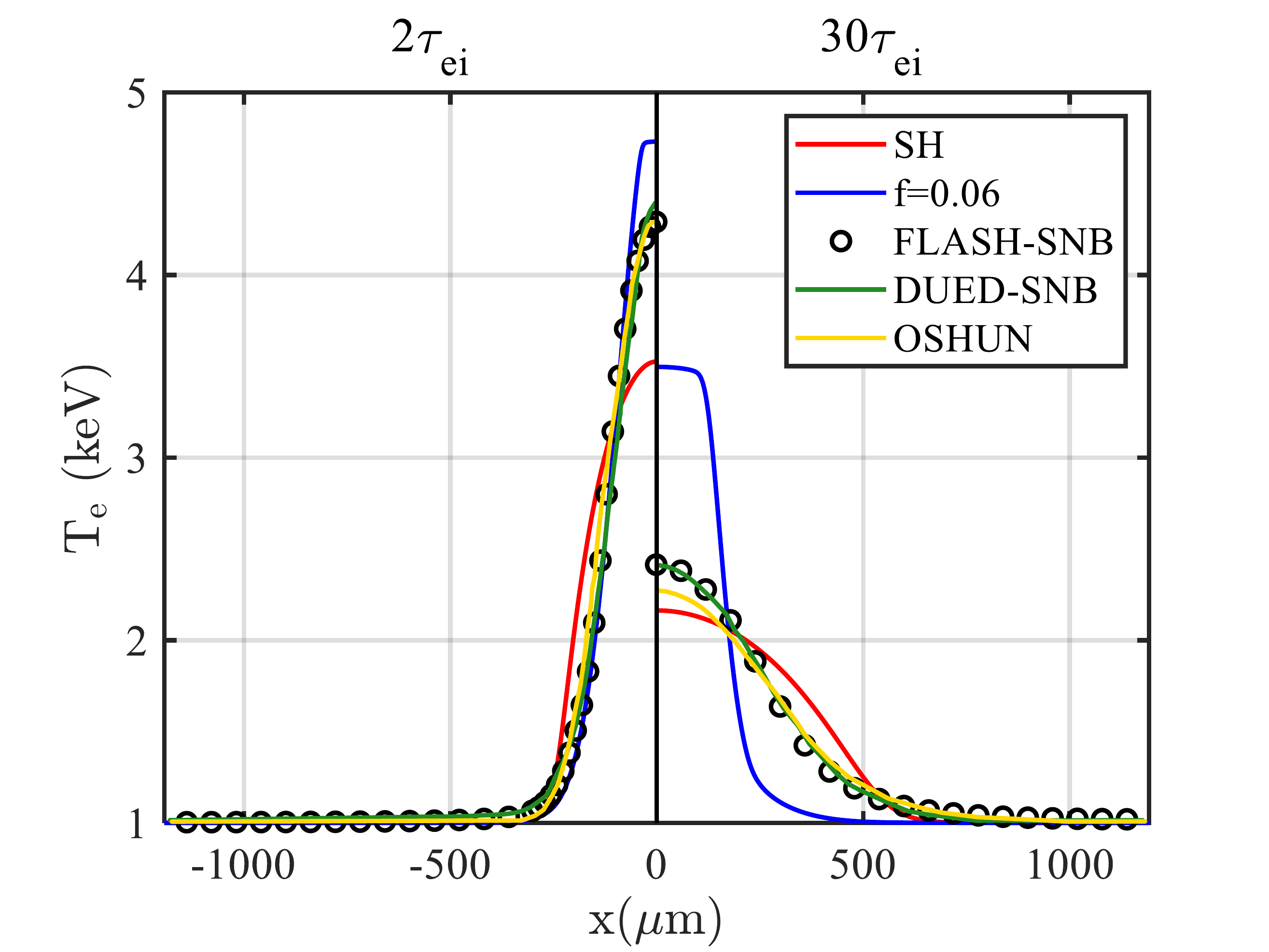}
	\caption{\label{fig2} Results of the hot spot relaxation test at two different time $2\tau_{ei}$ (left) and $30\tau_{ei}$ (right). The vertical coordinate is the electron temperature. Curves are shown for the SH model (red), the flux-limited model ($f = 0.06$, blue), our modified FLASH-SNB code (black dots). The DUED-SNB results (green) and the OSHUN results (yellow) given by Marocchino are also shown.}
\end{figure}

\begin{figure}[htbp]\centering
	\includegraphics[width=6cm]{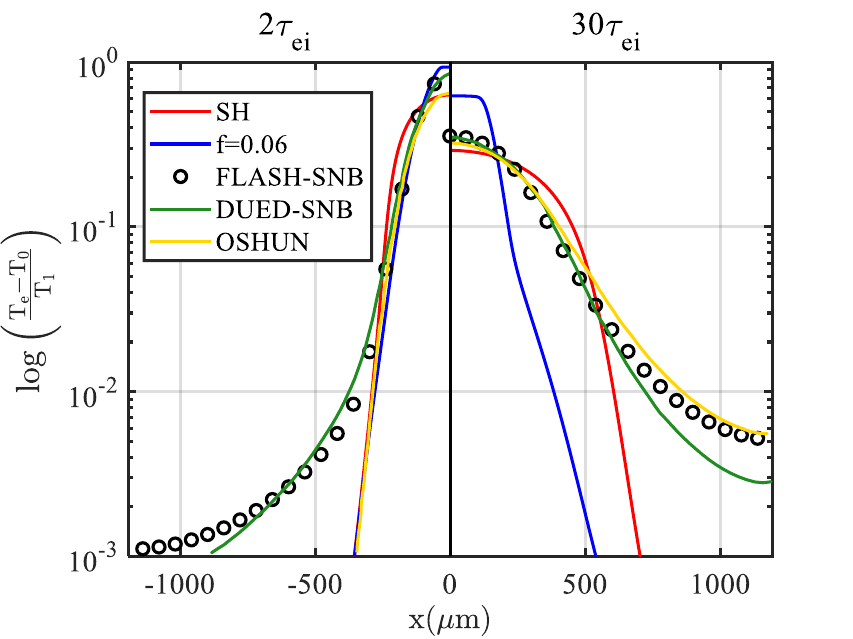}
	\caption{\label{fig3} Results of the hot spot relaxation test. The vertical coordinate is the dimensionless number $\log \left[ {\left( {{T_e} - {T_0}} \right)/{T_1}} \right]$. Curves are shown for the SH model (red), the flux-limited model ($f = 0.06$, blue), our modified FLASH-SNB code (black dots). The DUED-SNB results (green) and the OSHUN results (yellow) given by Marocchino are also shown.}
\end{figure}

The second example of calibration is the hot spot relaxation test. The initial temperature of the plasma has a Gaussian distribution. The state of the system at different time presents the kinetic and fluid-like behavior, respectively. We use the same parameters of Marocchino et al. \cite{marocchinoComparisonNonlocalHydrodynamic2013}, and show the results for the linear scale in Fig. \ref{fig2}, and logarithmic scale in Fig. \ref{fig3}. Figure \ref{fig2} shows that FLASH-SNB and DUED-SNB are consistent with each other both at $2\tau_{ei}$ ($x<0$) and at $30\tau_{ei}$ ($x>0$). Figure \ref{fig3} shows that only minor differences between that of DUED-SNB and FLASH-SNB in $x < -600\ \rm{\mu m}$ at $2\tau_{ei}$ and $x > 600\ \rm{\mu m}$ at $30\tau_{ei}$, with FLASH-SNB showing more preheat in the tail. The differences is arisen from the choice of the different dimensionless number r. In Marocchino's paper\cite{marocchinoComparisonNonlocalHydrodynamic2013} , the dimensionless number r = 16, while we chose r = 2 here. Therefore, the electron-ion mean free path here ($\lambda_g=2\sqrt2\left(\frac{E_g}{k_BT_e}\right)^2\lambda_{ei}$) is greater than their value ($\lambda_g=\left(\frac{E_g}{k_BT_e}\right)^2\lambda_{ei}$). This leads to the temperature diffusion in a broad region.

It is shown that the classical SH model is over-diffusive while the flux-limited model only reduces the heat flow above a given threshold, resulting in a flat top and a sharp temperature profile. The SNB model is significant differences with the classical SH model and the flux-limited model, indicating the necessity of using non-local models. In general, the results of FLASH-SNB match that of DUED-SNB and OSHUN. It can limit the heat flow correctly at steep temperatures, leading to a reasonable preheating, avoiding the artificial and non-physical setting of a flux-limiter.

\section{$\boldsymbol{2\omega}$ LASER ABLATING TARGETS}
\begin{figure}[htbp]
	\includegraphics[width=9cm,trim=30 0 30 0,clip]{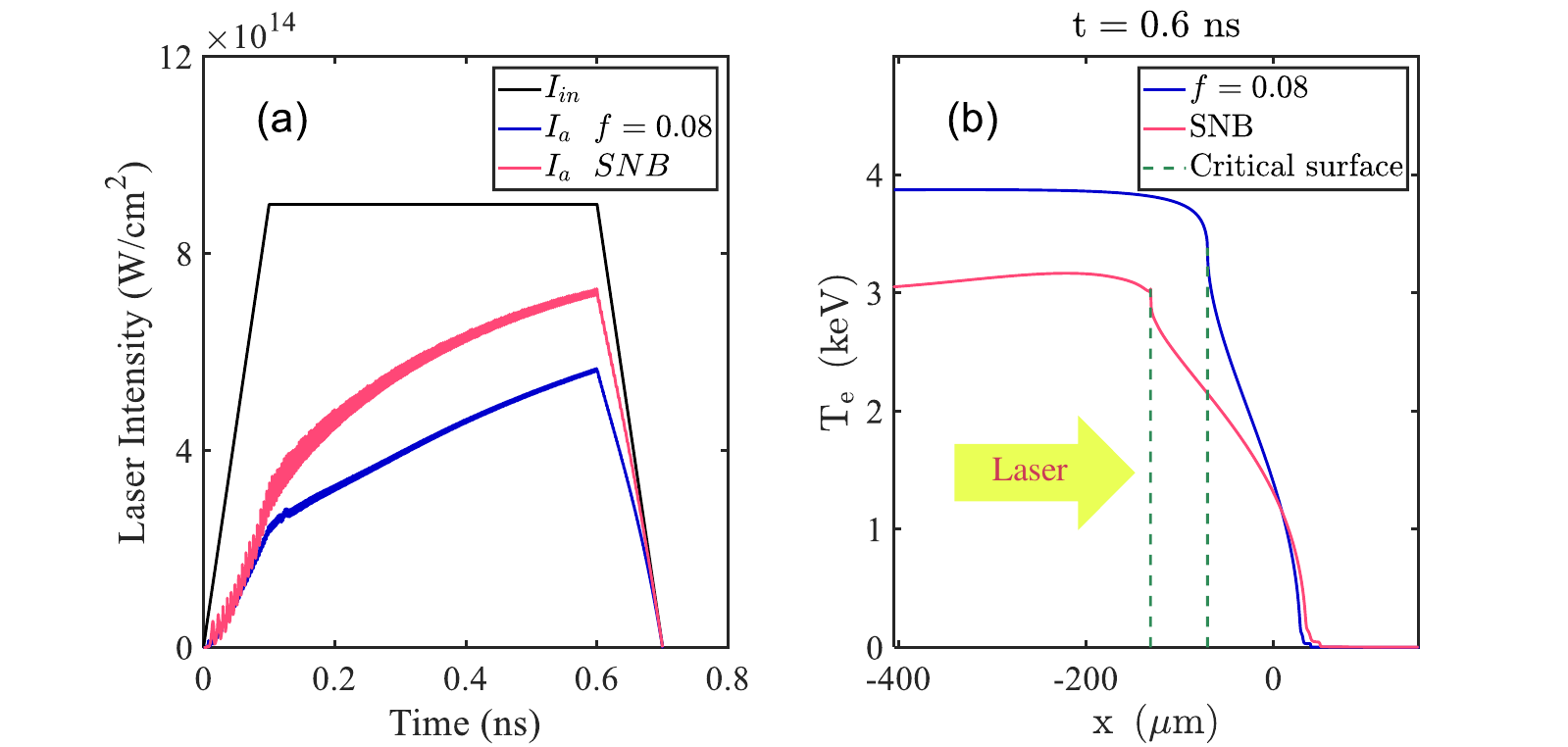}
	\caption{\label{fig4} (a) Absorbed irradiance of the $2\omega$ laser irradiated a CH target. The black curve is the input laser intensity with the peak plateau power of $9\times10^{14}\rm{W/cm^{2}}$. The red and black curves are the absorbed irradiance of the SNB model and the flux-limited model ($f$ = 0.08), respectively.(b)Temperature profile of the target along the laser propagation axis at $t=0.6$ ns. The critical surface temperature of the SNB model (red) is lower than that of the flux-limited model (blue).}
\end{figure}

\begin{figure}[htbp]
	\includegraphics[width=9cm,trim=30 0 30 0,clip]{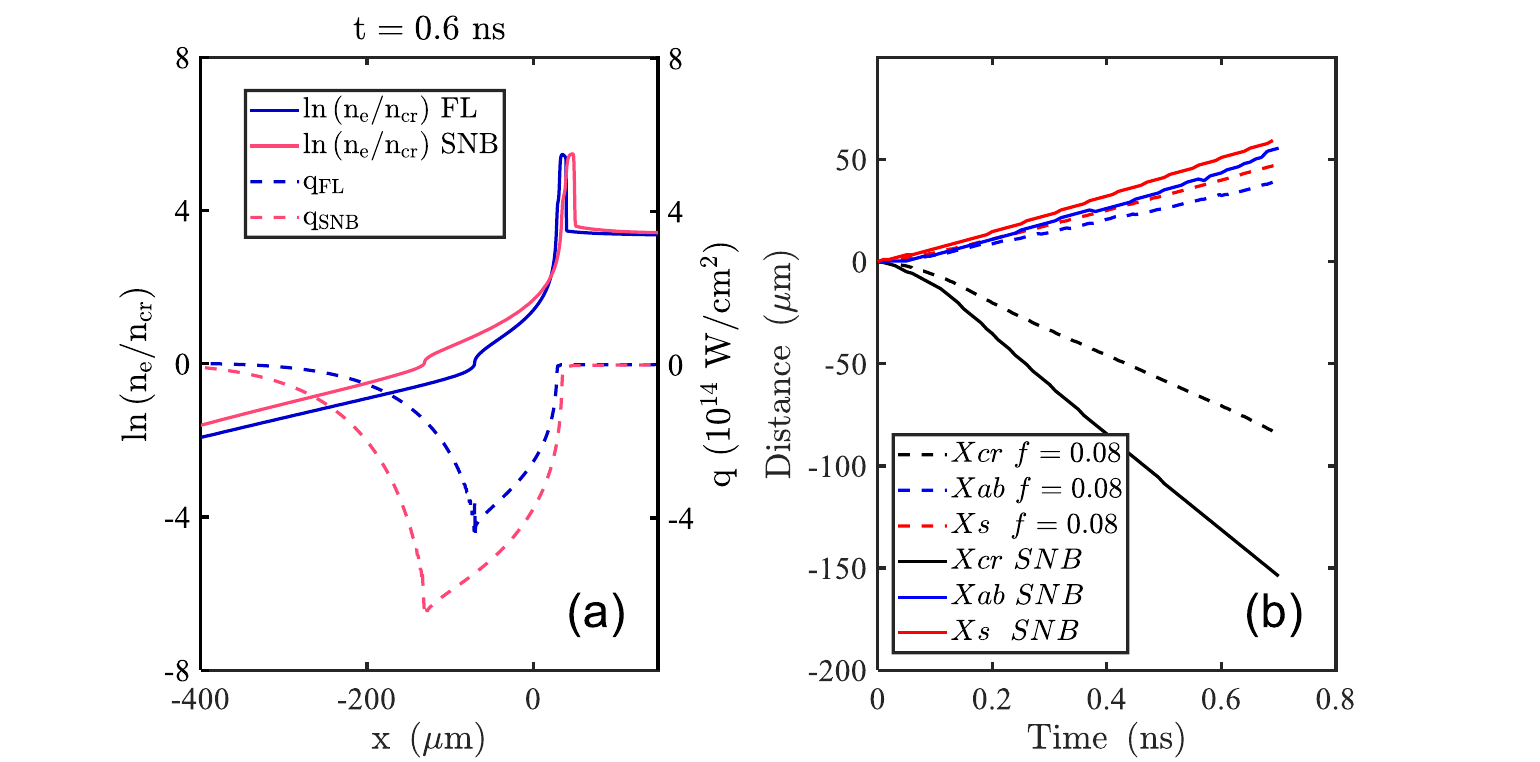}
	\caption{\label{fig5} (a) Density and heat flux profiles at $t$ = 0.6 ns for SNB model (red) and flux-limited model (blue). (b)Evolution of the position of the critical surface (black), ablation front (blue), and shock wave front (red) for the $2\omega$ laser irradiation of the CH target. The solid line is the SNB model and the dashed line is the flux-limited model ($f=0.08$).}
\end{figure}

We use a $2\omega$ laser to ablate the CH target with a peak laser intensity of $9\times10^{14}\ \rm{W/cm^{2}}$. The simulations are performed in the planar geometry and the thickness of the target is set to be semi-infinite. The temporal laser profile with a full width at half maximum of $0.6$ ns is shown in Fig. \ref{fig4}(a), which is referenced to the experimental parameters of Godsack et al.\cite{goldsackVariationMassAblation1982}. The equation of state parameters (EOS) are derived from QEOS \cite{moreNewQuotidianEquation1988} based on the Thomas-Fermi model, and the radiation opacity parameters are calculated by the SNOP program \cite{eidmannRadiationTransportAtomic1994} based on the mean atomic model. The energy groups of radiative transfer and non-local heat conduction of electrons are 20 and 30 groups, respectively. For most cases, the flux-limiter 0.08 matches the experimental results\cite{goncharovAblativeRichtmyerMeshkov2006}. Therefore, in the following studies, the flux limiter is set to 0.08. For the SNB model, the value of $\alpha$ of 0.01 has proved to be sufficiently accurate and the convergence in the low-density region ($\rho<10^{-4}\ \mathrm{g/cm^{3}}$) is ignored in order to avoid increasing computing resource by incorrect iterations at the boundary between the plasma and the vacuum. It can be seen that the absorbed irradiance of the SNB model is higher than that of the flux-limited model. The absorbed irradiance of 0.1$\sim$0.6 ns is integrated to obtain the average energy absorptivity, which is 63\%, while the absorptivity of the flux-limited model is 50\%. This is mainly due to the fact that the SNB model has a lower critical surface temperature, which induces a higher laser absorptivity of inverse Bremsstrahlung (IB) absorption. The temperature profile at 0.6 ns is presented in Fig. \ref{fig4}(b) and the profiles of density and heat flux are shown in Fig. \ref{fig5}(a) at the same moments. The temperatures at the critical surface of the two models are 3.6 keV and 3.0 keV, respectively. The higher temperature in coronal of the flux-limited model is due to the smaller flux-limiter, which leads to the heat flow being excessively limited. The laser energy deposited in the critical surface cannot be transferred quickly to the target, resulting in a higher coronal temperature. For the SNB model, the heat flow is not over-constrained, which results in a lower temperature in the corona as well as an increase the absorption of laser energy. 

We now discuss the heat conduction zone, i.e., the region from the critical surface to the ablation front. As shown in Fig. \ref{fig5}(b), the SNB model has a larger heat conduction zone, which is consistent with the phenomenon observed by Michel et al. \cite{michelMeasurementsConductionZoneLength2015}.The length of the heat conduction zone of the SNB model (209 $\mu$m) is about 65\% larger than that of the flux-limited model (128 $\mu$m). It is well known that the mass ablation rate and velocity at critical surface are dependent on the absorbed irradiance, i.e., $\dot m \propto {\left( {I_L^a} \right)^{1/3}}$, ${u_c} \propto {\left( {I_L^a} \right)^{1/3}}$. Since the SNB model has a higher energy absorptivity, the mass ablation rate and velocity at critical surface are higher than that of the flux-limited model. This leads to an increase in the length of the conduction zone, which helps to alleviate the hydrodynamic instability \cite{smalyukSystematicStudyRayleigh2008}. The shock wave front evolution is shown in Fig. \ref{fig5}. The shock speed of the SNB model is found to be about 18\% higher than that of the flux-limited model. The reason for the above phenomenon is also that non-local effects will lead to a higher mass ablation rate, inducing a higher ablation pressure and a more intense shock wave. 

\begin{figure}[htbp]
	\includegraphics[width=6cm]{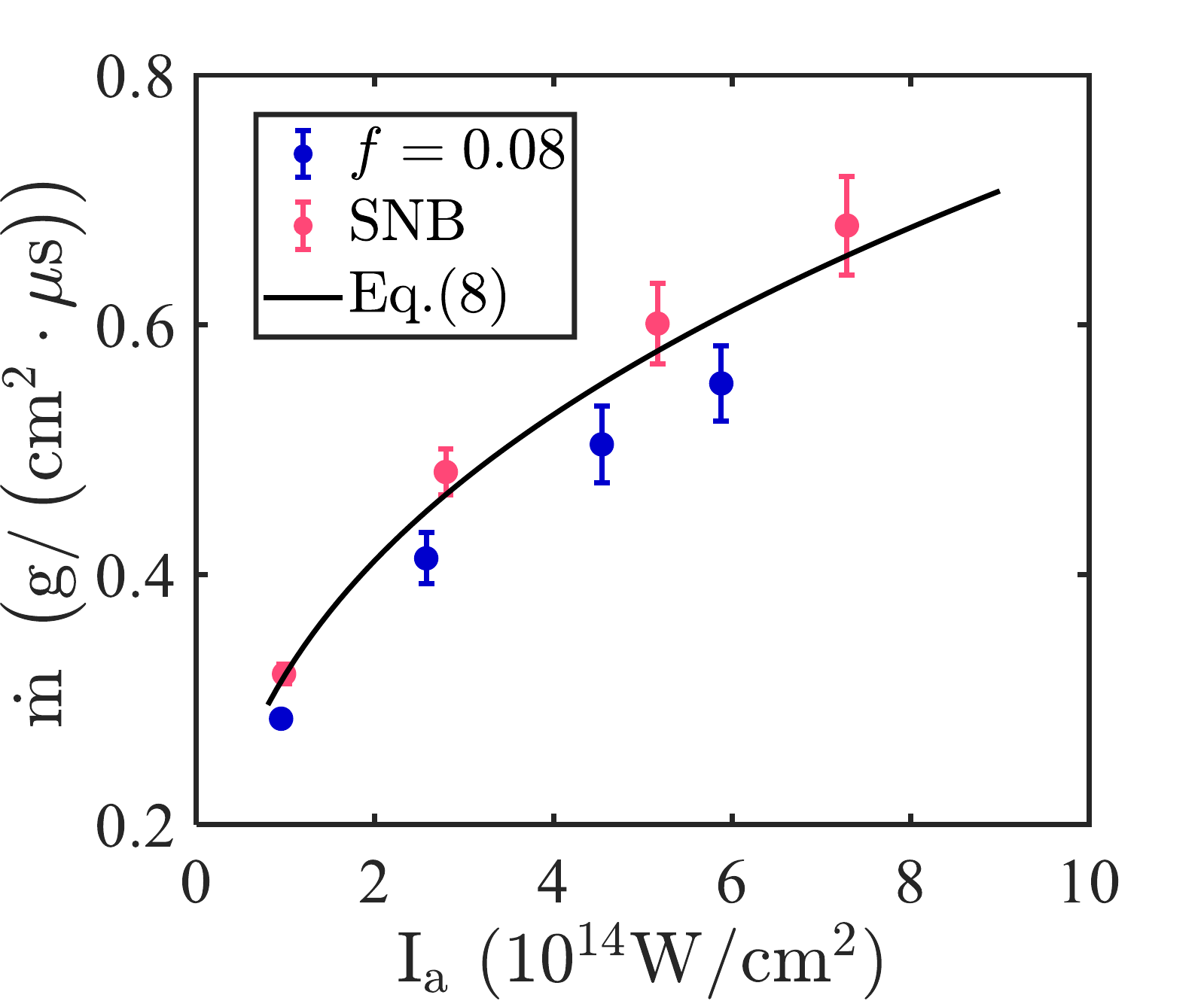}
	\caption{\label{fig6}Average mass ablation rates vs the absorbed irradiance. The red and blue dots are from the SNB non-local model and the flux-limited model, respectively, and the black curve is from the fitted experimental curve equations Eq. (\ref{Eq8}).}
\end{figure}

To investigate the influence of non-local transport under different laser intensities, we studied the mass ablation rate of the SNB model and the flux-limited model under different laser intensities, and the results are shown in the Fig. \ref{fig6}. The points in the Fig. \ref{fig6} denote the results of input laser peak power of $1\times10^{14}$, $3\times10^{14}$, $6\times10^{14}$, and $9\times10^{14}\ \rm{W/cm^{2}}$, respectively. It should be noted that since the laser energy absorption of the SNB model and the flux-limited model are different under the irradiation of the same laser, the horizontal coordinates of the points of the two models in Fig. \ref{fig6} do not overlap.

It can be seen that, in Fig. \ref{fig6}, the results of the SNB model (red dots) are in better agreement with the experimental results of Goldsack et al. \cite{goldsackVariationMassAblation1982} (black curve) than the flux-limited model. The experimentally fitted mass ablation rate is
\begin{equation} \label{Eq8}
	\dot m = 0.14{\left( {\frac{{{I_a}}}{{{{10}^{13}}{\rm{W/c}}{{\rm{m}}^{\rm{2}}}}}} \right)^{0.36}}\frac{{\rm{g}}}{{{\rm{c}}{{\rm{m}}^{\rm{2}}} \cdot {\rm{\mu s}}}},
\end{equation}
This illustrates that when the laser intensity is above $10^{14}\ \mathrm{W/cm^{2}}$, the non-local effects of the $2\omega$ laser are obvious and the value of flux-limiter of 0.08 is no longer applicable. With the increase of laser intensity, the mass ablation rate of the SNB model is between 1.12 and 1.23 times higher than that of the flux-limited model. The differences in absorbed irradiance between the two models are the main reason for the differences in mass ablation rate. The differences in laser energy absorption intensities between the two models are small for the peak power of $1\times10^{14}\ \rm{W/cm^{2}}$ and the energy absorption intensities of the SNB is only 3.5\% higher than that of the flux-limited model. It increases to 23.9\% for the peak power of $9\times10^{14}\ \rm{W/cm^{2}}$.

\section{INFLUENCE OF LASER FREQUENCY ON THE NON-LOCAL EFFECT}
\begin{figure}[htbp]
	\includegraphics[width=8.6cm,trim=15 0 15 0,clip]{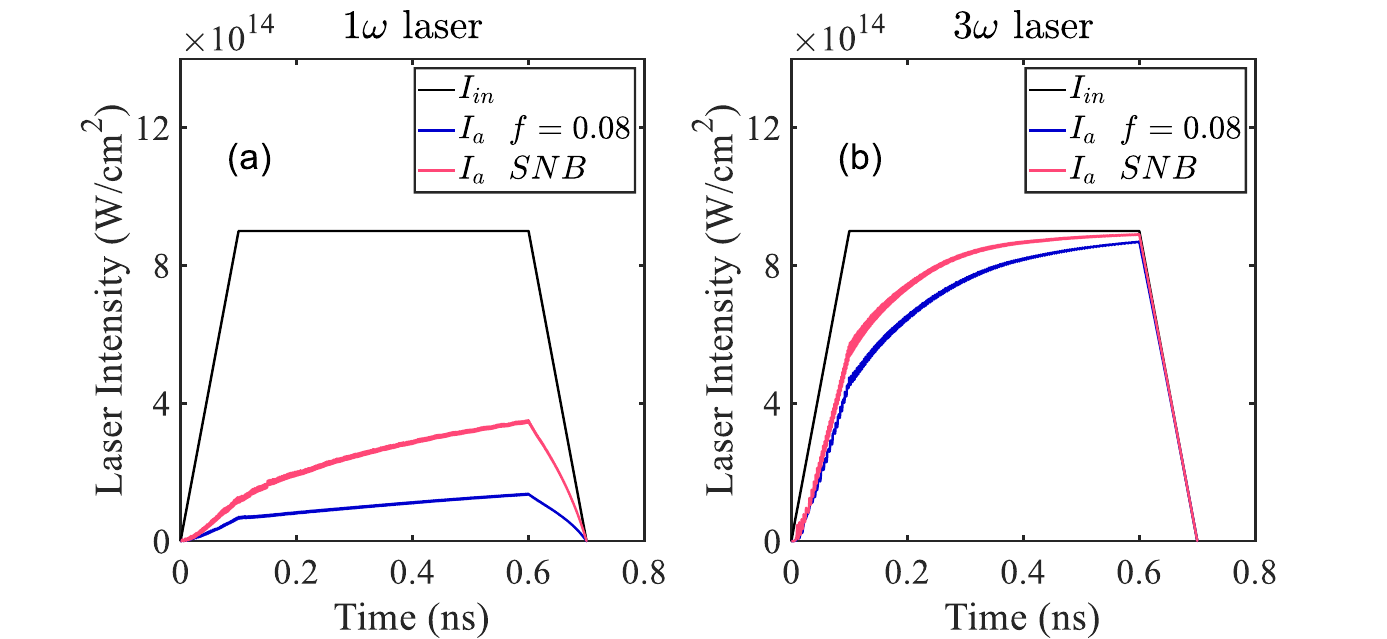}
	\caption{\label{fig7}The absorbed irradiance of the $1\omega$ laser (a) and $3\omega$ laser (b) irradiated CH target. The peak power is $9\times10^{14}\ \rm{W/cm^{2}}$. The black curve is the input laser intensity, and the blue and red curves represent the laser energy absorption intensity of the flux-limited model and the SNB model, respectively.}
\end{figure}

In order to understand the non-local transport effect completely, we study $1\omega$ and $3\omega$ lasers with a peak power of $9\times10^{14}\ \rm{W/cm^{2}}$ irradiating on the target. It is shown that, in Fig. \ref{fig7}(a), the energy absorption of the $1\omega$ laser is significantly smaller than that of the $2\omega$ laser. But the difference in laser energy absorption between the SNB model and the flux-limited model is larger than that of the $2\omega$ laser. We integrate the energy absorptivity from 0.1 to 0.6 ns and obtain an average laser energy absorptivity for the SNB model. The laser absorptivity is 28\%, which is in good agreement with the results given by Garban-Labaune \cite{garban-labauneEffectLaserWavelength1982}, while it is only 13\% for the flux-limited model. We performed the similar simulation for the $3\omega$ laser and the results are shown in Fig. \ref{fig7}(b). It is shown that the difference between the two models gets small for the same input power for the $3\omega$ laser. This indicates that the non-local effects are weak for the shorter laser wavelengths. It should be noted that the laser absorption may decrease due to neglect of the Langdon effect\cite{langdonNonlinearInverseBremsstrahlung1980}. However, the importance of the Langdon effect is dependent on $\frac{Zv_{osc}^2}{v_e^2}$, where Z is the ionization state, $v_{osc}$ is the peak electron oscillating velocity, and $v_{e}$ is the electron thermal velocity. The value of $\frac{Zv_{osc}^2}{v_e^2} $ is less than unity in all cases here, indicating that the Langdon effect on the laser energy absorption can be neglected here\cite{lucianiNonlocalHeatTransport1983}. We fit the scaling laws of mass ablation rate and ablation pressure. The mass ablation rate scales as $\dot m\sim I_a^{0.40}{\lambda ^{ - 1.19}}$ and the ablation pressure scales as ${P_a}\sim I_a^{0.75}{\lambda ^{ - 0.37}}$ in the case of SNB model, where $I_a$ is the absorbed intensity in units of $10^{15}\rm{W/cm^{2}}$ and $\lambda$ is the laser wavelength in micrometers. The mass ablation rate is $\dot m\sim I_a^{0.32}{\lambda ^{ - 1.34}}$ in Ref.~\onlinecite{goldsackVariationMassAblation1982}, and the result for the theoretical model\cite{fabbroPlanarLaserdrivenAblation1985,schmittImportanceLaserWavelength2023} is $\dot m\sim I_a^{1/3}{\lambda ^{ - 4/3}}$ and ${P_a}\sim I_a^{2/3}{\lambda ^{ - 2/3}}$. The scaling laws of the flux-limited model requires the constant selection for the flux limiter to get the correct results, and thus is not shown here.

\begin{figure}[htbp]
	\includegraphics[width=6cm,trim=15 0 15 0,clip]{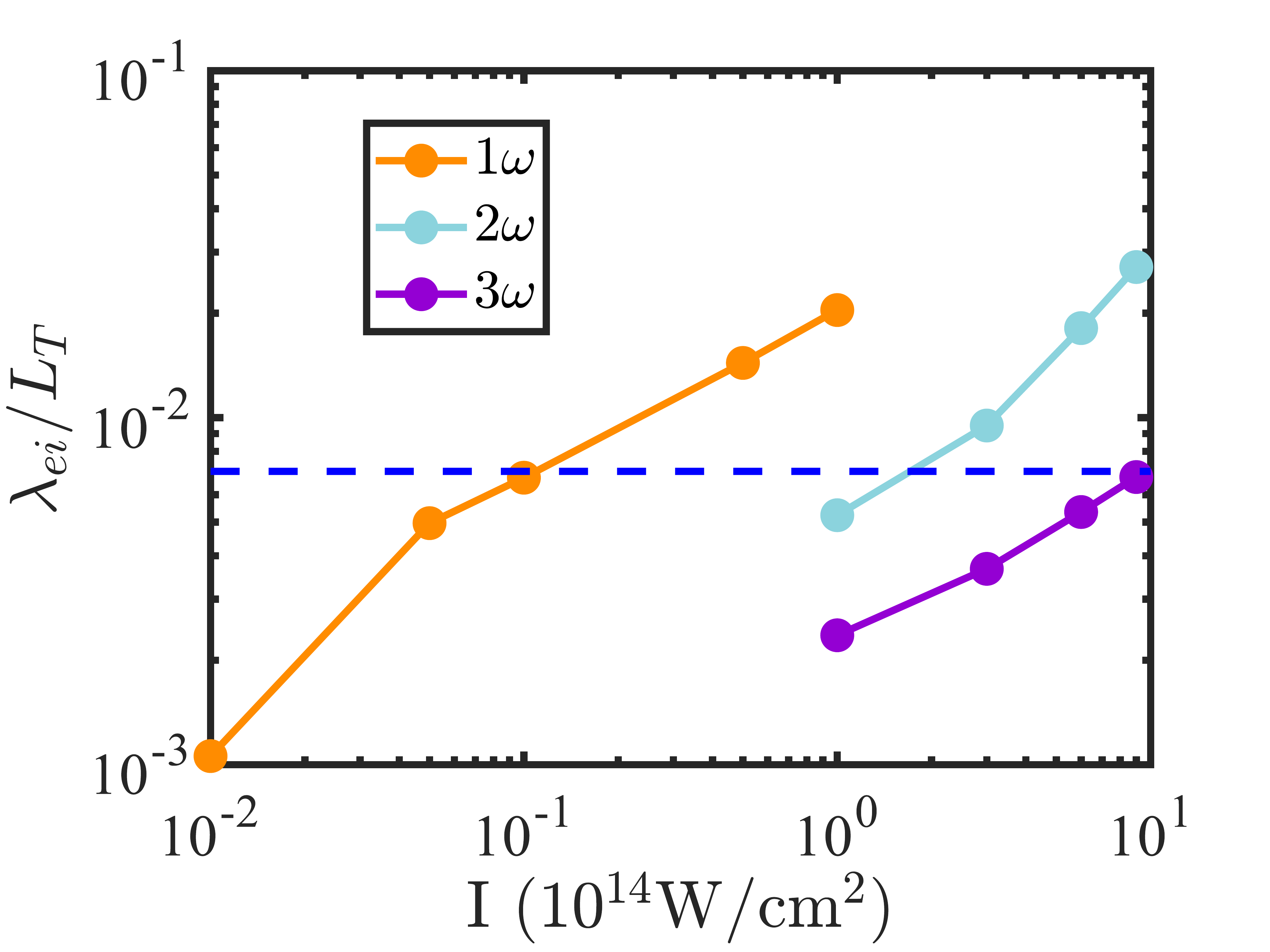}
	\caption{\label{fig8}Dependence of $\frac{\lambda_{ei}}{L_T}$ on laser intensity for different frequency lasers at $t=0.6$ ns.}
\end{figure}

To measure the effect of the non-local transport on the laser ablation, the Knudsen number around the critical surface is investigated in detail. The Knudsen number determines the extent that the electron distribution function deviates from the Maxwellian distribution. It depends on the electron temperature profile and the laser wavelength, i.e., $\frac{\lambda_e}{L_T}\propto\frac{T}{n_e}\frac{dT}{dx}\propto T\lambda_L^2\frac{dT}{dx}$. The Knudsen numbers at the critical density of the SNB model for different laser intensities and laser frequencies are shown in Fig. \ref{fig8}. As the laser intensity increases, the temperature profile becomes steeper and the scale length of the temperature decreases, thus the Knudsen number increases. The electron density around the critical surface decreases as the laser frequency increases. This increases the electron mean free path and the Knudsen number, and thus the non-local effect gets strong. We set Kn=0.007 as the critical point for separating local and non-local transport\cite{henchenMeasuringHeatFlux2019}. It is found that the thresholds that the non-local transport should be considered are $\sim 1\times10^{13}$, $\sim 2\times10^{14}$ and $\sim 1\times10^{15}\ \mathrm{W/cm^{2}}$ for $1\omega$, $2\omega$ and $3\omega$ lasers, respectively.

\begin{figure}[htbp]
	\includegraphics[width=8.6cm,trim=15 0 15 0,clip]{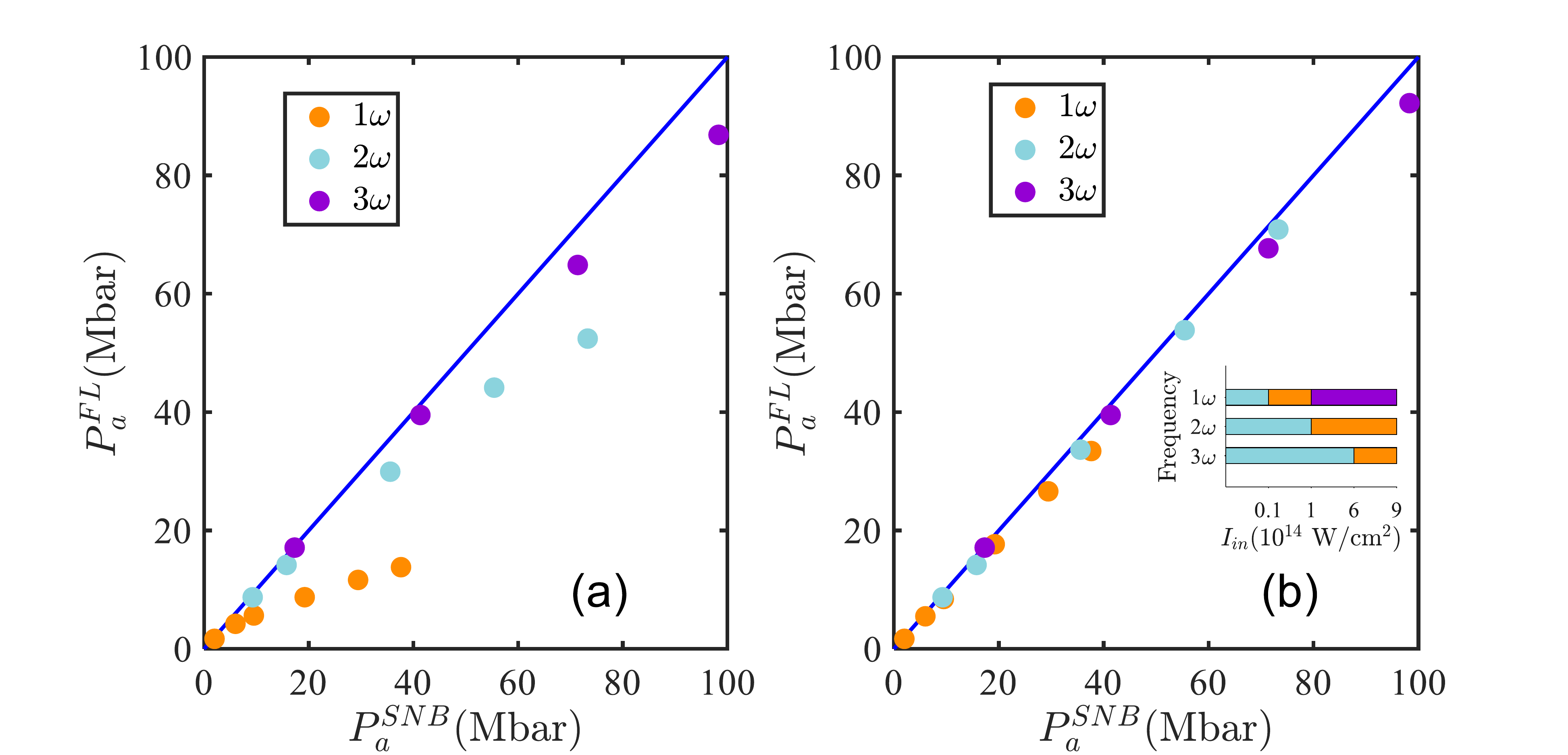}
	\caption{\label{fig9}Comparison of the ablation pressure between the SNB model and the flux-limited model. (a) SNB versus the flux-limited model with $f$ = 0.08. (b) SNB versus the flux-limited model with appropriate flux-limiters. The inset shows the applicable range of the flux-limiters $f$ = 0.08 (cyan histogram), $f$ = 0.14 (orange histogram) and $f$ = 0.20 (purple histogram).}
\end{figure}

Since the non-local effect varies with the laser intensity and frequency, the flux-limiter in different cases should be considered carefully. Figure \ref{fig9}(a) shows the ablation pressure for the SNB model and the flux-limited model (f = 0.08). The results for f = 0.08 deviate from the SNB model to different degrees with increasing laser intensity for different lasers. We thus chose proper flux-limiters to match the ablation pressure of the two models, and the results are shown in Fig. \ref{fig9}(b). The ablation pressures of the two models are highly consistent and the inset shows the use of the flux-limiters, where the cyan, orange, and purple histograms indicate $f$ = 0.08, 0.14, and 0.20, respectively. For a 1$\omega$ laser, $f$ = 0.14 with $I_{in}$ in the range of $10^{13}\sim10^{14}\mathrm{W/cm^{2}}$, and $f$ = 0.20 with $I_{in}$ in the range of $10^{14}\sim10^{15}\mathrm{W/cm^{2}}$. For a $2\omega$ laser, $f$ = 0.14 as $I_{in}$ is above $3\times10^{14}\mathrm{W/cm^{2}}$, while $f$ = 0.14 as $I_{in}$ is above $6\times10^{14}\mathrm{W/cm^{2}}$ for a $3\omega$ laser. That is, we obtain the applicable range of the flux-limiters for lasers with different frequencies and intensities. It should be noted that the results considering only the non-local effects in one dimension, and the effect of electron non-local transport on the laser ablation in two and three-dimensional will be investigated in our future work.

\section{CONCLUSIONS}
The improved SNB non-local electron transport model has been added into the radiation hydrodynamics code. The model for multigroup diffusion is based on the implicit algorithm proposed by Cao, and refers to the improved scheme of Sherlock for the mean free path $\lambda_{g}$ in multigroup diffusion. The non-local transport module is validated with two classical cases, which give consistent results. Then it is applied to investigate a $2\omega$ laser interaction with a CH target. It is shown that the non-local effect becomes significant when the laser intensity is higher than $1\times10^{14}\rm{W/cm^{2}}$. Compared with the standard flux-limited model, the SNB model has higher laser energy absorptivity due to the lower coronal plasma temperature, leading to a higher shock wave velocity, ablation velocity, and conduction zone length. It also has a higher mass ablation rate, which matches better with the experimental results compared to the flux-limited model. The laser intensity threshold for non-local effects is about $1\times10^{13}\ \rm{W/cm^{2}}$ for the laser of frequency $1\omega$, and this threshold increases to $\sim 2\times10^{14}\ \rm{W/cm^{2}}$ and $\sim 10^{15}\ \rm{W/cm^{2}}$ as the frequency increases to $2\omega$ and $3\omega$, respectively. The appropriate flux-limiters that should be employed in the flux-limited model for different lasers are also obtained. The results should be helpful for the laser-target ablation applications, especially for inertial confinement fusion.

\section*{DATA AVAILABILITY}
The data that support the findings of this study are available from the corresponding authors upon reasonable request.

\begin{acknowledgments}\suppressfloats
This work was supported by the National Natural Science Foundation of China (Grant Nos. 12175309, 11975308, 12005297, and 12275356), the Strategic Priority Research Program of Chinese Academy of Science (Grant No. XDA25050200 and XDA25010100), the State Key Laboratory of Laser Interaction with Matter (No. SKLLIM1908), X.H.Y. also acknowledges the financial support from Fund for NUDT Young Innovator Awards (No. 20180104).
\end{acknowledgments}



\bibliography{reference.bib}

\end{document}